\documentclass[doublespacing]{elsart}

\usepackage{amsmath}
\usepackage{graphicx}

\def\cal{\mathcal}

\begin{document}

\date{\today}

\begin{frontmatter}

\title
{Exciton Dephasing and Thermal Line Broadening in Molecular
Aggregates}

\author{
D. J. Heijs, V. A. Malyshev$^{1}$, J. Knoester$^*$}
\thanks{On leave from ``S.I. Vavilov State Optical Institute'',
199034 Saint-Petersburg, Russia.}

\address
{Institute for Theoretical Physics and  Material Science Centre,\\
University of Groningen, Nijenborgh 4, 9747 AG Groningen, The
Netherlands}

\corauth[corauthor]{ Corresponding author: J. Knoester, Institute
for Theoretical Physics and Materials Science Centre, University
of Groningen, Nijenborgh 4, 9747 AG Groningen, The Netherlands;
tel: +31-50-3634369; fax: +31-50-3634947; email: j.knoester@rug.nl
}

\begin{abstract}

Using a model of Frenkel excitons coupled to a bath of acoustic
phonons in the host medium, we study the temperature dependence of
the dephasing rates and homogeneous line width in linear molecular
aggregates. The model includes localization by disorder and
predicts a power-law thermal scaling of the effective homogeneous
line width. The theory gives excellent agreement with temperature
dependent absorption and hole-burning experiments on aggregates of
the dye pseudoisocyanine.

\end{abstract}


\journal{J. Luminescence}
\end{frontmatter}

\newpage

\section{Introduction}
    \label{intro}

Molecular J aggregates have fascinating optical properties caused
by the Frenkel exciton states arising from the strong excitation
transfer interactions in these systems. At low temperature, these
states give rise to narrow absorption lines (exchange
narrowing)~\cite{Knapp84} and ultra-fast collective spontaneous
emission (exciton superradiance)~\cite{deBoer90}. The temperature
dependence of the optical observables of molecular aggregates
shows many interesting features as well, and often differs
strongly from those of single molecules. For example, the
fluorescence lifetime typically increases with growing temperature
\cite{deBoer90,Fidder90,Potma98}, while the fluorescence Stokes
shift may exhibit an anomalous (non monotonous) temperature
dependence by showing a growth at low temperatures
\cite{Scheblykin01,Bednarz03}. Although the basic physics of these
behaviors is often well-understood, a consistent quantitative
explanation is mostly missing, due to the fact that the optical
dynamics in these systems results from the complicated interplay
between scattering of the excitons on disorder (localization) as
well as vibrations (dephasing). This also holds for the
explanation of the experimentally observed thermal growth of the
absorption line width and the homogeneous line width as measured
in photon echo and hole burning experiments. Several authors have
interpreted their measurements of this broadening as an activated
process caused by scattering on optical vibrations of the
aggregate \cite{Fidder90,Hirschmann89,Fidder95}, while others
concluded a power law thermal broadening, possibly arising from
excitons scattering on vibrations in the host medium
\cite{Renge97}. To add to the uncertainty, the surmised dephasing
mechanisms have not been correlated with the temperature
dependence of the fluorescence properties.

In this paper, we model the temperature dependent dephasing rates
of excitons in linear chains and use the results to analyze the
absorption line width and hole width as a function of temperature.
Our model includes scattering of the excitons on static disorder
as well as on acoustic vibrations in the host medium. The results
reveal a good agreement with  experiments on aggregates of the dye
pseudoisocyanine (PIC). We also comment on the successful
application of the model to the fluorescence properties of these
aggregates.

The outline of this paper is as follows. In section \ref{model},
we describe the model. The calculation of the scattering rates and
dephasing rates are addressed in section \ref{scattering}, where
also the calculation of the absorption spectrum and the hole
burning spectra are discussed. In section \ref{results}, we
present and discuss our numerical results, while in section
\ref{conclusions} we conclude.

\section{Model}
    \label{model}

Our model consists of a linear Frenkel chain of $N$ two-level
molecules ($n=1,\ldots,N$) with parallel transition dipoles. The
Hamiltonian of this system reads
\begin{equation}
\label{Hex}
    H
    = \sum_{n=1}^N \> \varepsilon_n |n\rangle \langle n|
    + \sum_{n,m=1}^N\> J_{nm} \> |n\rangle \langle m| \ ,
\end{equation}
where $|n\rangle$ denotes the state in which molecule $n$ is
excited and all others are in their ground state. The
$\varepsilon_n$ are the transition energies of the individual
chromophores, while the $J_{nm}=-J/|n-m|^3$ denote the
intermolecular excitation transfer interactions mediated by the
transition dipoles ($J>0$ is the interaction between nearest
neighbors). We account for disorder in the molecular energies,
caused by random solvents shifts, by taking the $\varepsilon_n$
from uncorrelated Gaussian distributions with standard deviation
$\sigma$ and average $\varepsilon_{0}$. The exciton eigenstates
(labeled $\nu$) follow from diagonalizing the $N \times N$ matrix
with the $\varepsilon_n$ as diagonal elements and the $J_{nm}$ as
off-diagonal ones. They have energy $E_{\nu}$, and site-amplitudes
$\varphi_{\nu n}$, i.e., the eigenstates read
\begin{equation}
    |\nu\rangle = \sum_{n=1}^N \> \varphi_{\nu n} |n\rangle.
\label{eigenstate}
\end{equation}
The disorder leads to localization of the excitons state on
segments of the chain; the optically dominant states, i.e., those
with most oscillator strength, occur in the low-energy tail of the
density of states (DOS), near the bottom of the disorder-free
exciton band.

The model includes on-site scattering of excitons on harmonic
vibrations (phonons) in the host medium through the
Hamiltonian~\cite{Davydov71}
\begin{equation}
\label{V1}
    V = \sum_{n=1}^N \> \sum_q V_{nq}
    |n\rangle \langle n|a_q + h.c.  \ ,
\end{equation}
where $a_q$ annihilates a phonon of quantum number (wave vector
and branch index) $q$, with energy $\omega_q$ (we set $\hbar=1$).
$V_{nq}$ indicates the strength of the linear exciton-phonon
coupling. In most J aggregates, in particular for those of the
prototypical PIC aggregates, the exciton-phonon interaction is
rather weak. This is evident from the narrowness of the absorption
line as well as the absence of a clear fluorescence Stokes shift.
As a consequence, the exciton-phonon interaction may be accounted
for in a perturbative way. This leads to transfer of population
from exciton state $\nu$ to state $\mu$ with a (scattering) rate
$W_{\mu \nu}$ that may be obtained from Fermi's Golden Rule. The
calculation of these rates is addressed in the next section.

\section{Scattering rates and spectra}
    \label{scattering}

Applying Fermi's Golden Rule to the exciton-phonon scattering
problem is straightforward. The result becomes particularly simple
for a disordered host (appropriate in most experimental
situations), in which the $V_{nq}$ are stochastic variables
uncorrelated for the various sites $n$. Referring to
Refs.~\cite{Bednarz03,Bednarz02} for details, we simply quote the
result:
\begin{equation}
\label{W}
    W_{\mu\nu} =   {\cal F}(|\omega_{\mu\nu}|)G(\omega_{\mu\nu})
    \sum_{n=1}^N \varphi_{\mu n }^2 \varphi_{\nu n}^2 \ .
\end{equation}
Here, $\omega_{\mu\nu} = E_\mu-E_\nu$ and ${\cal F}(\omega)=2 \pi
\sum_q |V_q|^2 \delta(\omega-\omega_q)$, the one-phonon spectral
density ($V_q$ characterizes the coupling of this mode to the
excitons). Furthermore, $G(\omega)=n(\omega)$ if $\omega>0$ and
$G(\omega)=1+n(-\omega)$ if $\omega<0$, with
$n(\omega)=[\exp(\omega/k_B T)-1]^{-1}$, the mean thermal
occupation number of a phonon mode of energy $\omega$.

The $W_{\mu \nu}$ may be used to study the intra-band relaxation
of excitons after photo-excitation by using them to set up a Pauli
master equation for the occupation probabilities of the exciton
states~\cite{Bednarz03,Shimizu01}. In addition, however, the
$W_{\mu \nu}$ govern the temperature dependent dephasing of the
excitons, and hence their homogeneous line widths. Specifically,
the thermal dephasing rate of the $\nu$th exciton state is given
by $\Gamma_{\nu} \equiv \frac{1}{2} \sum_\mu W_{\mu \nu}$ and
depends on temperature through the $n(\omega)$. The temperature
behavior of the dephasing rates depends sensitively on the form of
the spectral density. Here, we will assume that acoustic phonons
in the host medium dominate the scattering process. This leads to
${\cal F}(\omega) = W_0 (\omega/J)^3$, where the $\omega^3$
scaling derives from the fact that the phonon density of states
scales as $\omega^2$ (Debye behavior) and the $|V_q|^2$ scales
like $\omega$ for acoustic phonons. The remaining factor $W_0$ is
a free parameter in the model, which together with the disorder
strength $\sigma$, may be used to fit experimental data.

The above dephasing rates determine the absorption width of the
individual exciton states. As a consequence, the total absorption
spectrum reads:
\begin{equation}
\label{A}
    A(E) = \frac{1}{N} \left\langle \sum_\nu  \frac{F_\nu}{\pi} \,\,
    \frac{\Gamma_\nu + \gamma_\nu/2}{(E - E_\nu)^2 +
    (\Gamma_\nu + \gamma_\nu/2)^2}\right\rangle \ .
\end{equation}
where $F_\nu = (\sum_n \varphi_{\nu n})^2$ is the dimensionless
oscillator strength of the $\nu$th exciton state, $\gamma_\nu =
\gamma_0 F_\nu$ is its radiative rate ($\gamma_0$ is the emission
rate of a single chromophore), and the angular brackets denote
averaging over the static disorder in the $\varepsilon_n$. In the
next section we will report on numerical simulations of
Eq.~(\ref{A}). In addition, the above dephasing rates may be used
to calculate the shape and width of holes in hole burning
experiments. To do this, one restricts the summation over $\nu$ in
eq.~\ref{A} to those states that fall within the spectral interval
covered by the laser used to burn the hole. We will address
examples of this application in section \ref{results}.

\section{Results} \label{results}

It is useful to briefly consider the situation in the absence of
disorder ($\sigma=0$) with only nearest-neighbor interactions,
where the exciton states can be obtained analytically. The lowest
($\nu=1$) state in the band then carries more than 80\% of the
oscillator strength and it can be shown analytically that its
phonon-induced homogeneous width $\Gamma_1$ then scales with
temperature according to $T^{7/2}$. This power law directly
results from the $\omega^3$ dependence of the spectral density and
the $\omega^{1/2}$ dependence of the integrated DOS close to the
band bottom of one-dimensional systems. Inclusion of long-range
dipolar interactions slightly modifies the power law to
$T^{3.85}$, as follows from numerical calculations~\cite{Heijs05}.

In the presence of disorder, the dephasing rates of individual
exciton states vary considerably from one state to the other, due
to stochastic fluctuations in the energy differences and wave
function overlap summations that feature in
Eq.~\ref{W}~\cite{Malyshev01}. Especially at low temperature, the
width of the distribution of these rates may be of the same order
as their mean~\cite{Heijs05}. This makes it hard to identify a
typical homogeneous width. Instead, it is more useful to directly
simulate experimental observables.

\begin{figure}[ht]
\centerline{\includegraphics[width=0.7\columnwidth,clip]{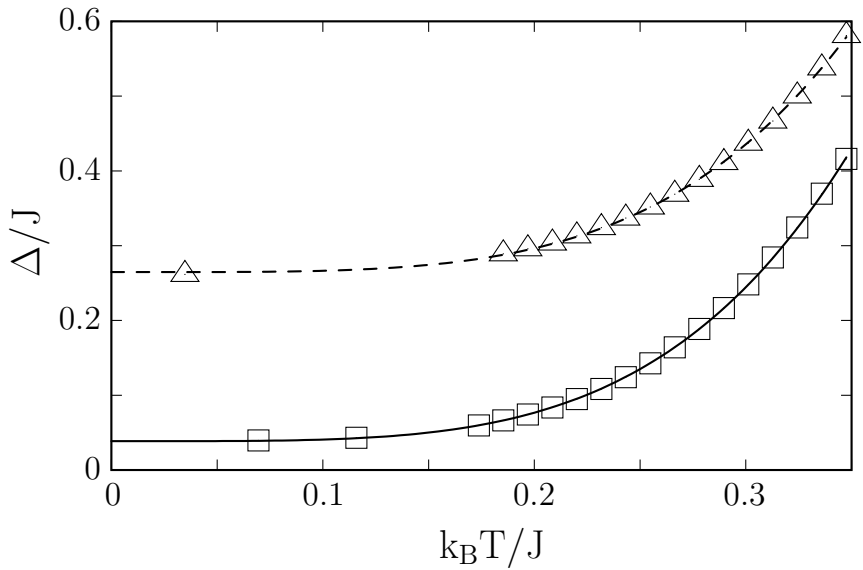}}
\caption{Calculated temperature-dependent width of the absorption
spectrum (symbols) and corresponding fits to Eq.~(\ref{scaling})
(lines) for two values of the disorder strength: $\sigma = 0.1J$
(solid) and 0.4J (dashed). The fit parameters $(a,p)$ take the
values (1.2, 4.2) and (1.2, 4.3), respectively. $W_0=25J$ and
$\gamma_0=1.5 \times 10^{-5}J$.} \label{figscaling}
\end{figure}

For chains of 250-1000 molecules, we simulated the linear
absorption spectrum according to Eq.~\ref{A} and extracted from it
the full width at half maximum $\Delta(T)$ as a function of
temperature. We have found that over an appreciable range of
$\sigma$ and $W_0$ values, $\Delta(T)$ follows an almost universal
power law given by
\begin{equation}
\label{scaling}
    \Delta(T) = \Delta(0) + aW_0(k_B T/J)^{p} \ ,
\end{equation}
where $a$ and $p$ depend only weakly on the parameters $\sigma$
and $W_0$. As an example, Fig.~\ref{figscaling} displays the
calculated $\Delta(T)$ values (symbols) for two sets of
parameters, together with the power-law fits. The (fit) parameters
are given in the caption. More generally, we have found that for
$0.05J < \sigma < 0.5J$, $p=4.2 \pm 0.2$, while $a=1.2 \pm 0.1$.
The scaling Eq.~\ref{scaling} implies that the total width may be
interpreted as the sum of an inhomogeneous width $\Delta(0)$ and a
dynamic contribution. We note that the power $p$ lies rather close
to the value of 3.85 obtained in the absence of disorder. The
reason is that the states to which most of the population is
scattered lie higher in the exciton band, where  the DOS is not
affected much by disorder.

Interestingly, Renge and Wild~\cite{Renge97} found that the total
J-bandwidth $\Delta(T)$ of aggregates of PIC, with counter ions
chloride as well as fluoride, over a wide temperature range (from
10 K to 300 K) follows a power-law scaling as in
Eq.~(\ref{scaling}). Although the power reported by these authors
($p=3.4$) is smaller than the ones we obtained  above, from direct
comparison to the experimental data we have found that our model
yields excellent fits to the experimental data over the entire
temperature range, both for the shape and the width of the J-band.
This also holds for the J-bandwidth of PIC-Br (bromide), measured
between 1.5 K and 180 K~\cite{Fidder90}. Details will be published
elsewhere~\cite{Heijs05a}.

\begin{figure}[ht]
\centerline{\includegraphics[width=0.7\columnwidth,clip]{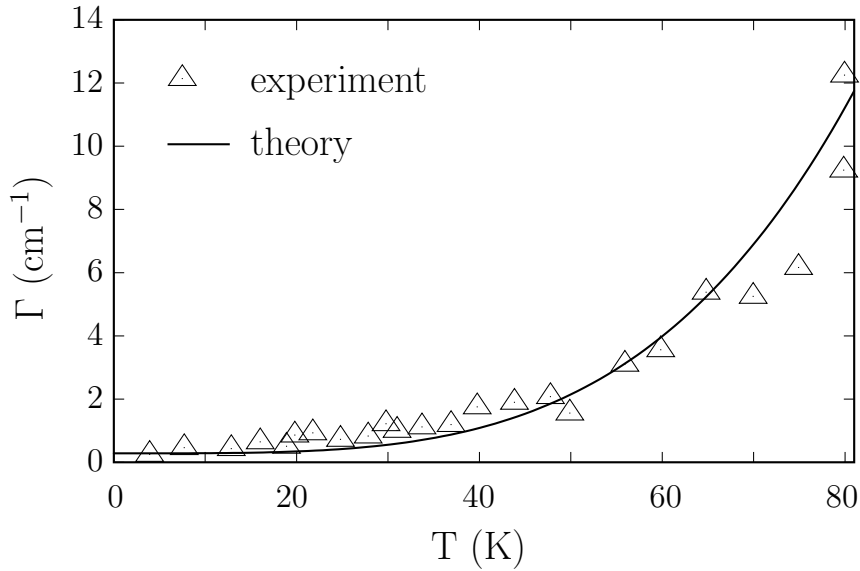}}
\caption{Hole width $\Gamma$ as a function of temperature measured
for aggregates of PIC-I after pumping in the center of the J-band
(triangles)~\cite{Hirschmann89}. The solid line is our fit; model
parameters are discussed in the text. } \label{holeburning}
\end{figure}

We have also calculated hole-burning spectra, according to the
method described at the end of section~\ref{scattering}. We
present here a fit to the results reported by Hirschmann and
Friedrich~\cite{Hirschmann89}. Using the hole-burning technique,
they measured the homogeneous width of the exciton states in the
center of the J-band for PIC-I (iodide) over the temperature range
350 mK to 80 K. Their data for the hole width $\Gamma$ are
reproduced as triangles in Fig.~\ref{holeburning}. The solid line
shows our fit to these data, obtained by simulating disordered
chains of $N=250$ molecules. The resonant interaction strength and
the monomer radiative rate were chosen at the accepted values of
$J=600$ cm$^{-1}$ and $\gamma_0 = 1.5 \times 10^{-5}J = 2.7 \times
10^8$ s$^{-1}$, respectively. Thus, the only free parameters were
the disorder strength $\sigma$ and scattering strength $W_0$.
First, we fixed the value of $\sigma$ by fitting the
low-temperature (4 K) absorption spectrum, where the homogeneous
broadening may be neglected. This yielded $\sigma = 0.21 J$. Next,
$W_0$ was adjusted such that the measured growth of the hole width
was reproduced in an optimal way. Thus, we found $W_0=180J$.

We observe from Fig.~\ref{holeburning} that our model yields a
good fit to the measurements. We have also tried to fit the
absorption and hole-burning data using a spectral density that (on
average) scales with $\omega$ according to a power different from
$3$ and found that is not possible to obtain a good fit. This
suggests that scattering on acoustic phonons of the host dominates
the dynamics of excitons in PIC aggregates. This conclusion
differs markedly from previous interpretations, where the optical
vibrations of the aggregate itself were considered as the dominant
scattering agents, leading to sums of activated temperature
dependencies~\cite{Hirschmann89,Fidder95}. In order to describe
the strong increase of the line width for large temperatures, such
a fit requires both a high activation energy (several hundreds of
cm$^{-1}$) and a large pre-exponential factor. The pre-factor is
directly related to the exciton-vibration coupling constant. A
simple perturbative treatment shows that the thus estimated
coupling constant should be so large (few thousand
cm$^{-1}$'s)~\cite{Heijs05} that a strong polaron effect is to be
expected. This is inconsistent with the generally accepted
excitonic nature of the optical excitations in cyanine aggregates.

\begin{figure}[ht]
\centerline{\includegraphics[width=
0.7\columnwidth,clip]{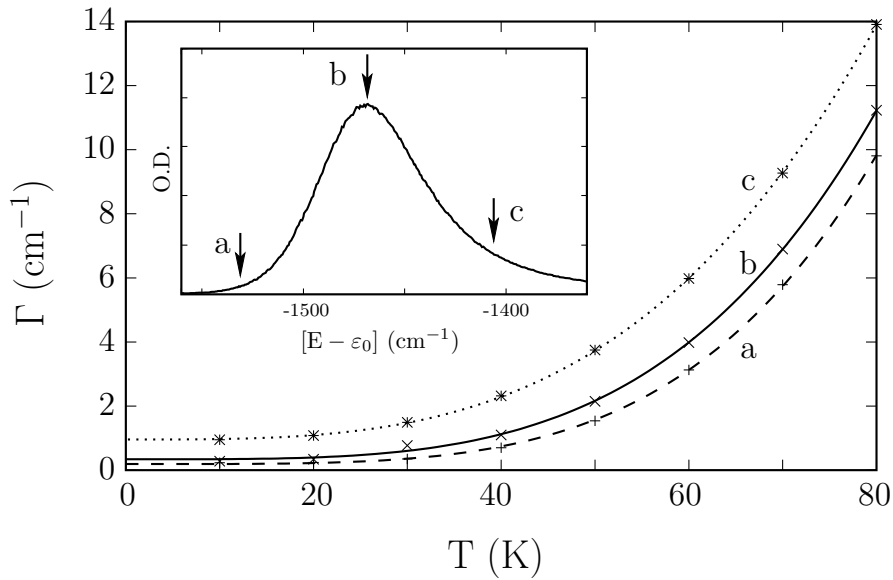}} \caption{Theoretical hole
widths $\Gamma$ as a function of temperature for burning
frequencies in the red wing (dashed, a), the center (solid, b) and
the blue wing (dotted, c) of the absorption spectrum. In the inset
the arrows indicate the burning frequencies in the absorption
spectrum of the aggregate at a temperature of 30 K. Model
parameters as in Fig.~\ref{holeburning}. } \label{threeholes}
\end{figure}

Using the same parameters as for the for of the PIC-I hole-burning
data in Fig.~\ref{holeburning}, we have also calculated the
temperature dependence of the hole width after burning in the red
and blue wings of the absorption line, respectively. Specifically,
we used burning frequencies taken $\Delta(T)$ to the red of the
peak of the absorption band and to the blue of the peak. The
results are given in Fig.~\ref{threeholes} together with the
results when burning at the peak position. It is clearly seen that
the hole width increases with increasing burning frequency. At low
temperatures, this is to be expected:  the dephasing of  the
excitons in the red wing (curve {\it a} in Fig.~\ref{threeholes})
is determined solely by radiative decay, whereas for the states at
higher energy (curves {\it b} and  {\it c} in
Fig.~\ref{threeholes}) also relaxation to lower-energy exciton
states contributes to the dephasing rate. The latter contributes
grows with growing exciton energy. With increasing temperature,
the hole width increases for all three burning frequencies,
because downward as well as upward scattering on phonons becomes
more prominent. It turns out that for all cases, the hole width
scales with temperature according to a power-like
Eq.~(\ref{scaling}), where the power $p$ depends on the burning
frequency (4.1 for the red, 3.8 for the peak, and 3.3 for the blue
side). At higher temperatures, the hole widths for the three
burning frequencies tend towards each other, because upward
scattering in the exciton band then dominates the dephasing rates.
We note that the frequency dependence of the dephasing rate within
J-band has also been observed using photon-echo experiments, see,
e.g., Ref.~\cite{Lampoura02}.

\section{Concluding remarks} \label{conclusions}

In summary, we have studied the temperature dependent dephasing
rate of exciton states in disordered Frenkel chains, caused by
scattering on acoustic phonons that are characterized by a
Debye-like ($\propto \omega^3$) spectral density. We have found a
power-law temperature dependence for the absorption line width as
well as the hole width obtained through hole burning. The model
yields good quantitative explanations to absorption and
hole-burning experiments on aggregates of the dye pseudoisocyanine
with various counter-ions.

We finally note that our model does not account for two-phonon
scattering processes; in particular we neglected pure dephasing.
We have investigated such processes and found that for acoustic
phonons their contribution to the dynamic line width of the
excitons scales according to $T^s$, with $s$ in the range
7-8.5~\cite{Heijs05}, which does not agree with the experimental
findings. Apparently the contribution of such processes is not
noticeable in aggregates of PIC. This may partly result from the
fact that the pure-dephasing rate is suppressed by a factor that
is equal to the delocalization size of the exciton states
\cite{Heijs05}, i.e., typically by a factor of 50.

\end{document}